\documentclass[reprint, amsmath, amssymb,aps, prx,longbibliography,superscriptaddress,floatfix]{revtex4-1}

\usepackage{amsmath,amssymb,bm,graphicx}
\usepackage{graphics}
\usepackage{float} 
\usepackage[breaklinks=true,colorlinks,citecolor=blue,linkcolor=blue,urlcolor=blue]{hyperref}
\def\be{\begin{equation}}
\def\ee{\end{equation}}
\def \bea{\begin{eqnarray}}
\def \eea{\end{eqnarray}}

\def \formula{$\mathrm{MoSi}_{2}{\mathrm{Z}}_{4}$}
\def \formulaAs{$\mathrm{MoSi}_{2}{\mathrm{As}}_{4}$}
\def \formulaN{$\mathrm{MoSi}_{2}{\mathrm{N}}_{4}$}
\def \formulaP{$\mathrm{MoSi}_{2}{\mathrm{P}}_{4}$}
\def \hsp{\hspace{2cm}}

\begin{document}

\title{Room temperature electron-hole liquid phase in  
monolayerMoSi$_2$Z$_4$ (Z = pinctogen)}
\author{Pushpendra Yadav}
\email{pyadav@iitk.ac.in}
\affiliation{Department of Physics, Indian Institute of Technology Kanpur, Kanpur-208016, India}
\author{K. V. Adarsh}
\email{adarsh@iiserb.ac.in}
\affiliation{Department of Physics, Indian Institute of Science Education and Research Bhopal, Bhopal 462066, India}
\author{Amit Agarwal}
\email{amitag@iitk.ac.in}
\affiliation{Department of Physics, Indian Institute of Technology Kanpur, Kanpur-208016, India}

\begin{abstract}
Photo-excited electrons and holes in insulators, above a critical density and below a critical temperature, can condense to form an electron-hole liquid (EHL) phase. However, observing the EHL phase at room temperature is extremely challenging. Here, we introduce the monolayer {\formula} (Z = N, As, P) series of compounds as a promising platform for observing the EHL phase at room temperature. The higher impact of the Coulomb interactions in two dimensions helps these monolayers support the EHL phase with an increased EHL binding energy and transition temperature, along with strongly bound excitons.  Our findings motivate further exploration of the {\formula} monolayers for realizing the EHL phase at high temperatures to harness collective phenomena for optoelectronic applications. 
\end{abstract}
\maketitle

\begin{figure}
	\centering
	\includegraphics[width =\linewidth]{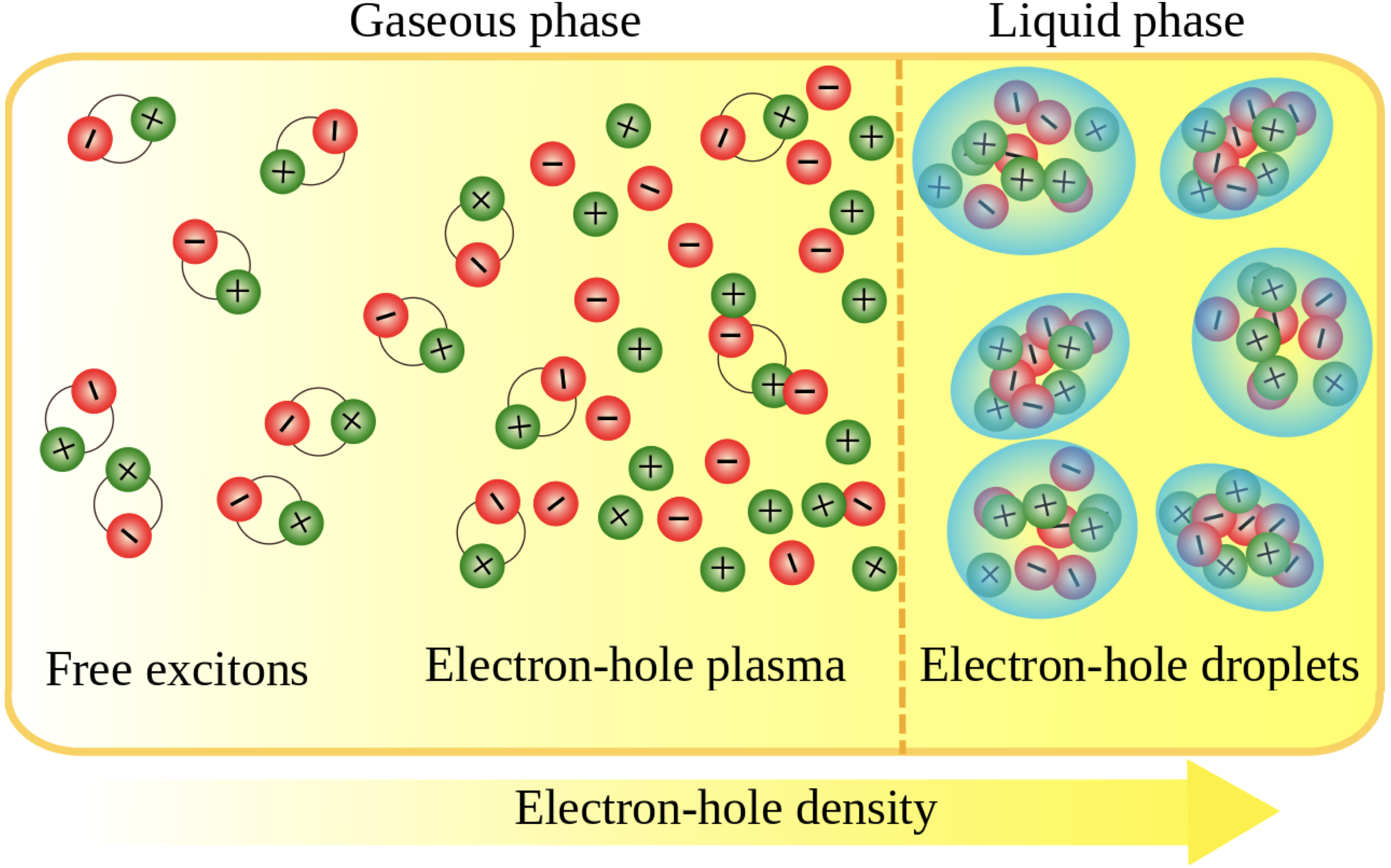}
	\caption {Schematic showing the formation of the electron-hole liquid from photo-excited electrons and holes. The free excitons dissociate on increasing photo-excited carrier density and form the electron-hole plasma state. In both of these phases, the constituents interact weekly with each other and can be treated as a gaseous state. Further increase in the exciton density leads to the formation of electron-hole droplets or the EHL phase, with the particles interacting collectively.} 
	\label{Fig1}
\end{figure}

The electron-hole liquid (EHL) state is an exciting example of a phase transition in the non-equilibrium regime~\cite{keldysh1968,Jeffries,Morita2022,Luo2020,asnin1969,MOS2-EHP-Bataller2019,MoS2-EHL-EXP-ACS-Nano-2019-Yu}. It arises from the condensation of electrons and holes in a photo-excited system at high carrier densities~\cite{Arp2019,Keldysh-Silin-75,AlmandHunter2014,Jia2021,wolfe-75, Beni-Rice-76}. This condensation transforms the electron-hole pairs into a metallic, degenerate Fermi liquid state. At low photo-excited carrier densities and temperatures, excitons are formed, and they interact weakly, forming a non-interacting free-exciton gas~\cite{Rice1978-book,walker-87}. As carrier densities increase, the pairwise Coulomb attraction between electrons and holes is screened, and excitons dissociate into electrons and holes, resulting in an electron-hole plasma (EHP) state~\cite{MOS2-EHP-Bataller2019}. Further increase in the carrier density strengthen the collective interaction between electrons and holes, leading to their condensation into droplets (see Fig.~\ref{Fig1}) with a rich phase diagram~\cite{Brinkman-72,Brinkman-Rice-73, Beni-Rice-78, landau, rice-conference,Shah-1977,Simon-1992,Vashishta-PRL-1974,Berciaud2019}. 

Unfortunately, the EHL phase generally occurs at extremely low temperatures. 
This is dictated by the EHL binding energy, which is typically one-tenth of the exciton binding energy. As a consequence, the EHL transition temperature ($T_c$) satisfies, $k_BT_c < 0.1 E_{\rm ex}$~\cite{landau,MoS2-nano-lett}, where $E_{\rm ex}$ is the exciton binding energy. 
The exciton binding energies in 3D semiconductors range from $0.1-0.001$ eV resulting in $T_c < 20$ K~\cite{Beni-Rice-78, Vashishta-PRL-1974}. However, their 2D counterparts have exciton binding energies of 100s of meV \cite{SOC-effect-MoS2-PRL,spin-orbit-MoS2-PRB,TMD-BE,amit-exciton-1, Pekh2020,pekh2021phase-1, MoS2-nano-lett, MoS2-EHL-EXP-ACS-Nano-2019-Yu}, owing to the reduced dielectric screening of the Coulomb interaction. For example, monolayer MoS$_2$ has an exciton binding energy of  $E_{\rm ex} \approx 0.6$ eV. Due to this, 2D semiconductors offer an ideal platform to observe the EHL phase at room temperature. 
Recent photoluminescence experiments and theoretical calculations have shown the possibility of room temperature EHL phase in MoS$_2$~\cite{MoS2-EHL-EXP-ACS-Nano-2019-Yu,MoS2-nano-lett,AB-Ray-EHL-MoS2}. 
This has instigated our search for other 2D systems supporting the room-temperature EHL phase, which can open new avenues for exploring non-equilibrium phase transitions without the limitation of cryogenics.

\begin{figure*}
	\centering
	\includegraphics[width =0.9\textwidth]{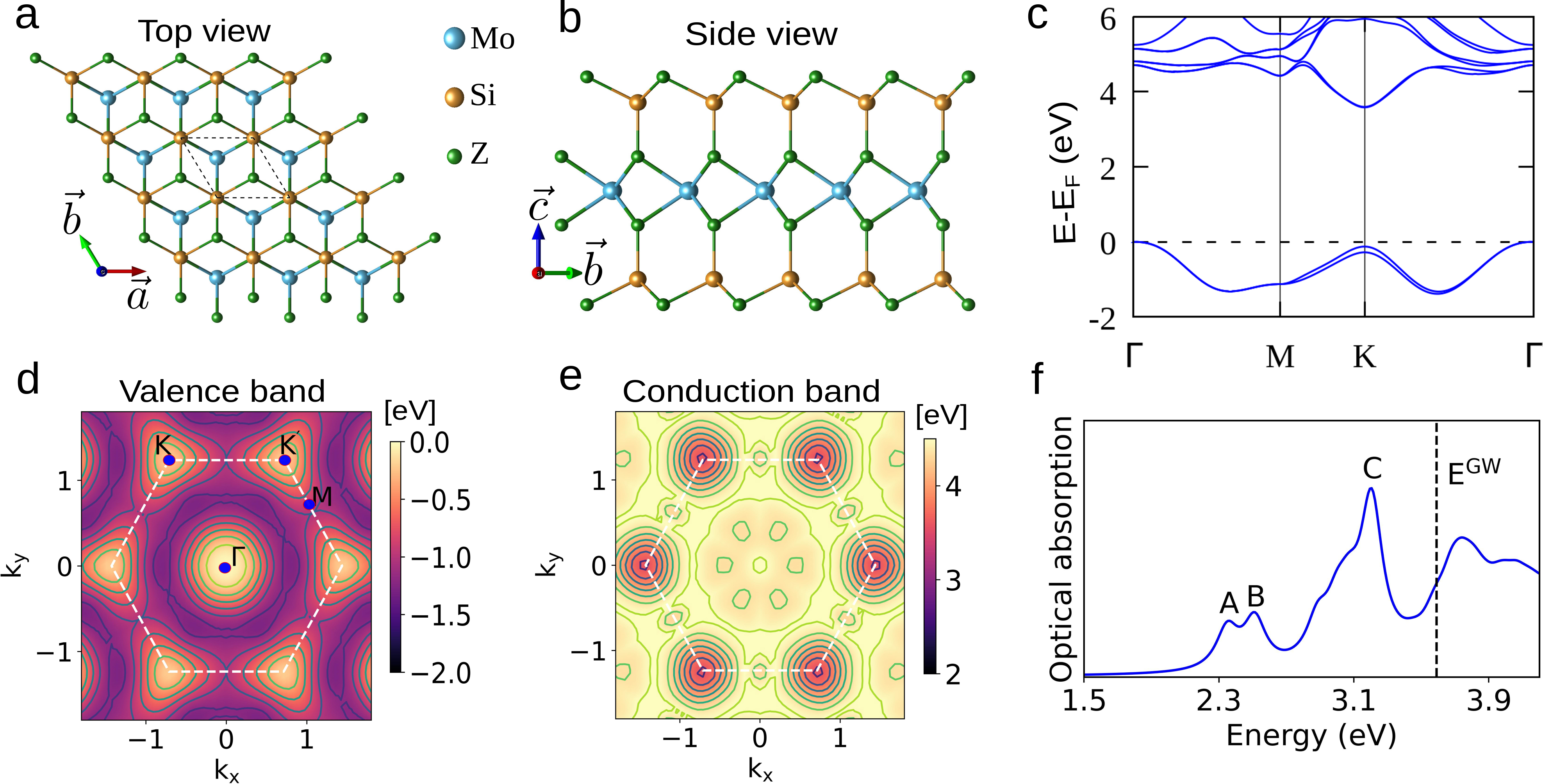}
	\caption { \textbf{a}) The top and \textbf{b}) side view of the  crystal structure of monolayer {\formula} (Z= N, As, P). The  Z-Si-Z-Mo-Z-Si-Z arrangement of atoms along the $c$-axis can be clearly seen in panel \textbf{b}). \textbf{c}) The electronic band structure for {\formulaN}. Our GW calculations indicate {\formulaN} to be an insulator with a bandgap of 3.58 eV. The band dispersion of \textbf{d}) the highest occupied valence band and \textbf{e}) the lowest unoccupied conduction band in the full BZ. Both the valence and conduction bands are degenerate at the $\mathrm{K}$ and $\mathrm{K^\prime}$ valleys of the 2D BZ. \textbf{f}) The optical absorption spectrum was calculated using the Bethe-Salpeter equation for monolayer {\formulaN}. Several prominent excitonic absorption peaks (A, B, and C) emerge within the quasiparticle bandgap. \label{Fig2}}
\end{figure*}

Here, we predict that monolayer {\formula} series (Z = N, P, or As) can host the EHL phase for temperatures above room temperature. 
This is facilitated by the strongly bound excitons in the {\formula} series having binding energies of up to 1000 meV~\cite{Hong670, pushpendra_exciton22, KONG2022, Wu2022, Sun2022}.
We calculate the ground state energy and the phase diagram of the EHL phase in the {\formula} series, taking into account the kinetic, exchange, and correlation energy of the electron-hole pairs. 
Our phase diagram predicts that monolayer {\formulaN} can sustain the EHL phase below a critical temperature of $T_c \sim 415$ K and for photo-excited carrier densities higher than $n_c \sim 10^{11}$ cm$^{-2}$.
Our findings open new avenues for exploring the non-equilibrium quantum many-body EHL state in monolayer {\formula} series for potential quantum technology and high-power laser applications. 

%
%

\begin{table*}[t]
	\def \hsp{\hspace{10cm}}
	\caption{The lattice parameter $a$ in Angstrom (\AA) for the monolayers of {\formula} (Z = N, As, P) and their electronic bandgap (E$_g$) in electron-Volt (eV), calculated within GW method. The effective masses of electrons ($m^*_e$) and holes ($m^*_h$) are listed in terms of electron mass ($m_e$). The number of electron and hole valleys are represented as $\nu_e$ and $\nu_h$ for the lowest unoccupied conduction band minima and the highest occupied valence band maxima at the $K/K^{\prime}$ of the 2D hexagonal Brillouin zone.}	
	\vspace{0.15 cm}
	\begin{tabular}{c c c c c c c c c }
		\hline \hline \vspace{1 mm}
		Compound  & $a$ (\AA) & E$_{g}$ (eV) & $m^*_e/m_e$   & $m^*_h/m_h$ &  $\nu_e$  & $\nu_h$ \\
		\hline
		\formulaN  & 2.909  & 3.58  & 0.407{\cite{Valley-pseudospin-PRB}} & 0.554{\cite{Valley-pseudospin-PRB}} & 2 & 2 \\
		\formulaAs & 3.621  & 1.70  & 0.499{\cite{nanotech-MoSi2As4}} & 0.419{\cite{nanotech-MoSi2As4}} & 2 & 2 \\
		\formulaP  & 3.471  & 1.74 & 0.325{\cite{nanotech-MoSi2As4}} & 0.393{\cite{nanotech-MoSi2As4}} & 2 & 2 \\
		\hline
	\end{tabular}
	\label{table:1}
\end{table*}

The structure of monolayer {\formula} is hexagonal and it is shown in Fig.~\ref{Fig2} (a) and Fig.~\ref{Fig2} (b). In our \textit{ab-initio} calculations, we used previously reported lattice parameters for the MoSi$_2$Z$_4$ series from Hong et al. \cite{Hong670}, which are also summarized in Table~\ref{table:1}. See the Methods section for the details of the density functional theory calculations and the quantum many body perturbation theory calculations (GW and Bethe-Salpeter). 
Our electronic structure calculations reveal that monolayer {\formulaN} is an indirect bandgap semiconductor with a bandgap of $3.58$ eV [see Fig.~\ref{Fig2} (c)]. In contrast, {\formulaP} and 
{\formulaAs} are direct bandgap semiconductors with bandgaps of $1.74$ eV and $1.70$ eV at the $K$ and $K^{\prime}$ points of the Brillouin zone (BZ), respectively. See Fig. S1 of the Supplementary material (SM) \footnote{Supplementary material has details about i) the GW bandstructure and absorption spectrum for {\formulaAs} and {\formulaP}, ii) fluence dependence of the bandgap renormalization and iii) the impact of the dielectric constant and the effective layer thickness on the phase diagram.}. 

The breakdown of inversion symmetry in all three hexagonal {\formula} monolayers makes them gapped. The strong spin-orbit coupling (SOC) of the Mo-$d$ orbitals in these materials lifts the valence band degeneracy at the $K$ and the $K^{\prime}$ points. The energy dispersion calculation for full 2D BZ reveals that valence (and conduction) band maxima (minima) possess the same energy at the K and K$^\prime$ valleys [see Fig.~\ref{Fig2} (d) and Fig.~\ref{Fig2} (e)].    
 Using the QP energies of electrons and holes, we construct the Bethe-Salpeter equation (BSE) to incorporate the electron-hole interactions. We obtain the optical absorption spectrum~\cite{BSE-1-Strinati, BSE-2-Strinati, BSE-3-Onida} by numerically solving the BSE to extract the imaginary part of the dielectric function. All our quantum many-body perturbation theory calculations are done using the implementation in the YAMBO package~\cite{yambo20091392,yambo2019}. 
 
 The optical absorption for the monolayer {\formulaN} is shown in Fig.~\ref{Fig2} (f),  with the GW band gap marked by a vertical line. The absorption peak in the bandgap highlights the presence of excitonic bound states. The lowest energy excitons peaks (A and B in Fig.~\ref{Fig2}) have a binding energy of 1.35 eV. The calculated optical bandgap and the binding energies are consistent with the previously reported experimental~\cite{Hong670} and theoretical~\cite{pushpendra_exciton22, Wu2022, Sun2022, KONG2022} values. The optical absorption spectrum of {\formulaAs} and {\formulaP} is shown in Fig.~S1 of the SM \cite{Note1}. The strongly bound excitons in these monolayers  open up the possibility of  an EHL phase at room temperature~\cite{MoS2-nano-lett}. To explore this further, we calculate the phase diagram of the photo-excited electrons and hole in these monolayers, starting with its ground state energy.


The total ground state energy of a system with interacting electrons and holes comprises of kinetic, exchange, and correlation energies. The total kinetic energy of  a 2D electron-hole system can be approximated as,  
\begin{equation}
E_{\text{kin}} = \sum_{i=e,h}\nu_i \sigma_i \sum_{k<k_F^i}\frac{\hbar^2 k^2}{2m_i}~ = ~\sum_{i=e,h} \frac{1}{2}E_F^{i}~.
\label{eq1}
\end{equation}
Here, $i = e/h$ represents electrons/holes, $\sigma_i$ ($\nu_i$) is the spin (valley) degeneracy of the bands, and $k_F^i$ is the corresponding Fermi wavevector. In Eq.~\eqref{eq1}, the Fermi wavevector for each species is given by $k_F^{i} = k_F/\sqrt{\nu_{i}}$ where, $k_F=(2\pi n)^{1/2}$ and $n$ represents electron-hole pair density. The corresponding Fermi energy is specified by $E^{i}_F = \hbar^2(k_F^{i})^2/2m_{i}$. 
The valley structure of the {\formula} series of monolayers can be seen from panels (d) and (e) of Fig.~\ref{Fig2}. We find that for all three compounds, the optically relevant band extrema occurs at the $K$ and $K'$ points of the BZ. This implies $\nu_e = \nu_h = 2$ for all the three studied monolayers. The effective masses of electrons and holes near these band extrema are obtained from the $ab-initio$ band structure calculations for the monolayer {\formula} and tabulated in Table~\ref{table:1}.

The impact of coulomb interactions can be split into the exchange and the correlation contributions. The exchange contribution is captured within the first-order perturbation theory by calculating the expectation value of the Coulomb interaction Hamiltonian using the multi-particle eigenstates of the non-interacting Hamiltonian~\cite{Brinkman-Rice-73, Giuliani-Giovanni-Vignale,Bergersen_1975}. 
For a generic Coulomb potential specified by $V_k$ (in the momentum space), it has the following form~\cite{MoS2-nano-lett,Pekh2020,pekh2021phase-1},  
\begin{equation}
E_{\text{exch}} = -\sum_{i=e,h}\frac{\nu_i \sigma_i}{2L^2} \sum_{k,q<k_{F,i}}V_{k-q}~,
\label{eq2}
\end{equation}
where $L^2$ is the area of the 2D system.

For a freestanding 2D system of zero thickness, the unscreened Coulomb potential is given by $V_k = 2 \pi e^2/k$. However, most of the 2D crystalline systems have a finite width and can be encapsulated on both sides by substrate of different dielectric constants. These effects are captured by the Keldysh potential~\cite{keldysh1986} which has the form, 
\bea
V_k = \frac{2\pi e^2}{\epsilon'k(1+r_0k)}~.
\label{eq3}
\eea
Here, $\epsilon^{\prime}$=($\epsilon_1$ + $\epsilon_2$)/2 with $\epsilon_1$ ($\epsilon_2$) being the dielectric constant of the top (bottom) substrate and $r_0$ is the effective thickness of the 2D system~\cite{Angel-Rubio2011_r0,MacDonald_r0}. The Keldysh potential is known to be more accurate for calculating the exciton binding energy of the monolayer transition metal dichalcogenides~\cite{MoS2-nano-lett}. 
It also captures the change in the impact of the Coulomb interaction across dimensional crossover. The Keldysh potential reduces to the unscreened 2D Coulomb potential in the $r_0 \to 0$ limit. For large $r_0$ values, it mimics the 3D Coulomb potential with $V_k \propto 1/k^2$ (Fig.~S3 of the SM \cite{Note1}). 
The effective thickness $r_0$ in monolayer {\formula} series is calculated using the relation $r_0 \propto d/\epsilon$ where $d$ is the layer thickness and $\epsilon$ is the dielectric constant of the material. For monolayer MoS$_2$, we have $d$ = 3.1 \AA, and $r_0/d = 14.16$ \cite{MacDonald_r0,Angel-Rubio2011_r0,MoS2-nano-lett}. 
Using this $r_0/d$ value for monolayer {\formulaN},  {\formulaAs}, and {\formulaP} we estimate their $r_0$ to be 99.15 \AA, 140.65 {\AA}, and 132.12 \AA, respectively.

\begin{figure}
	\centering
	\includegraphics[width =\linewidth]{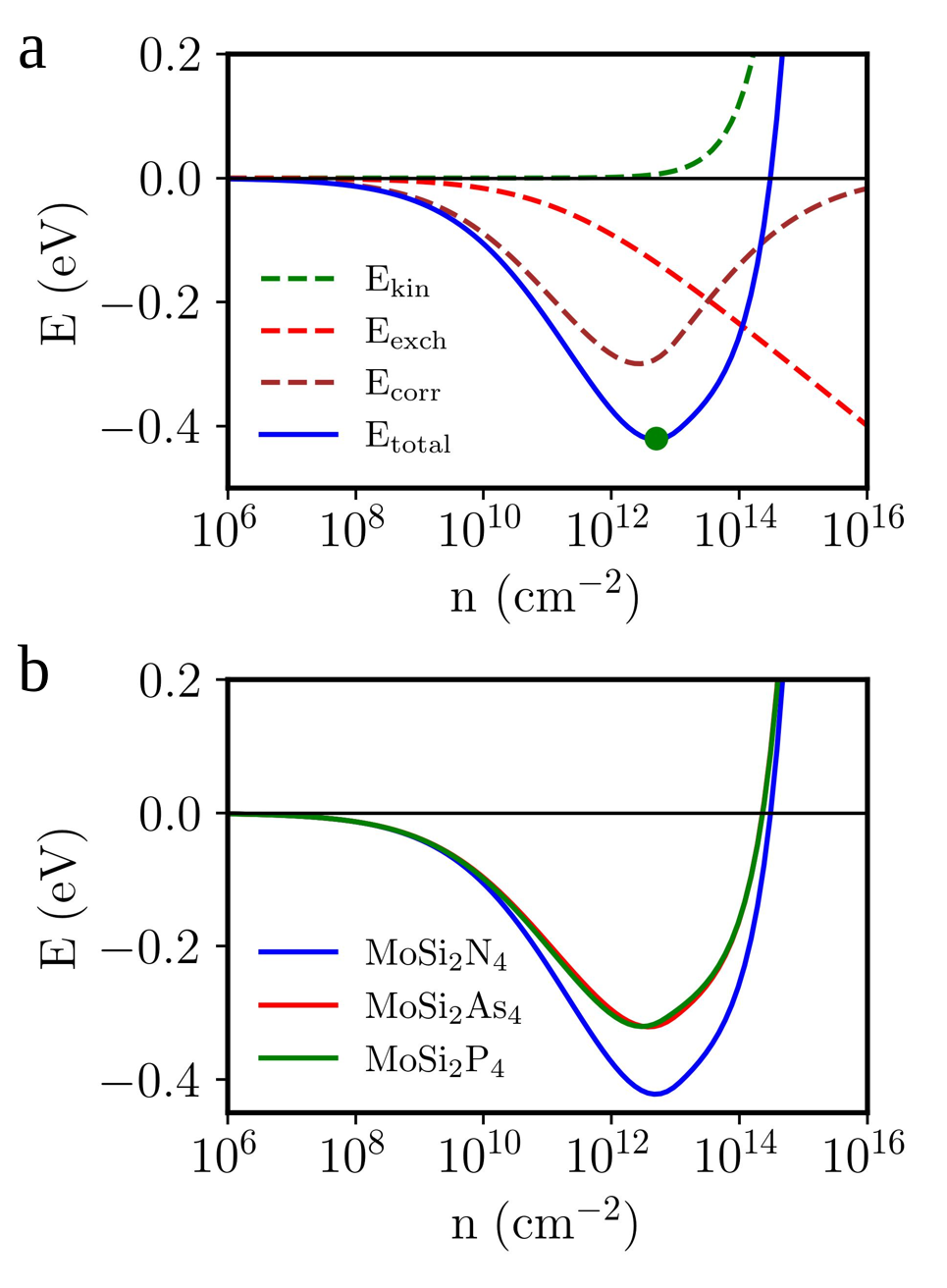}
	\caption {The total ground state energy as a function of photo-excited carrier density $n$ for the monolayer {\formula} series. \textbf{a}) The kinetic, exchange,  correlation, and total ground state energy (dashed green, red, brown, and solid blue curve, respectively) for monolayer {\formulaN}. The kinetic energy is prominent in the high-density limit, while the correlation energy dominates in the low-density regime. \textbf{b}) The total ground state energy ofx the three monolayers. The photo-excited electrons and holes in monolayer {\formulaN} have the lowest ground state energy. 
	\label{Fig3}}
\end{figure}

In contrast to the exchange energy for unscreened Coulomb potential~\cite{Pekh2020,pekh2021phase-1}, the analytical form of the exchange energy for the Keldysh potential (Eq.~\ref{eq3}) is not known. Therefore,  we calculate it numerically. The dependence of the exchange energy for monolayer {\formulaN} is shown 
by the red dashed curve in Fig.~\ref{Fig3} (a). 
As a check of our calculations, we numerically reproduce the known exchange energy for unscreened Coulomb interactions in 2D. 

\begin{figure*}
	\centering
	\includegraphics[width =.7\linewidth]{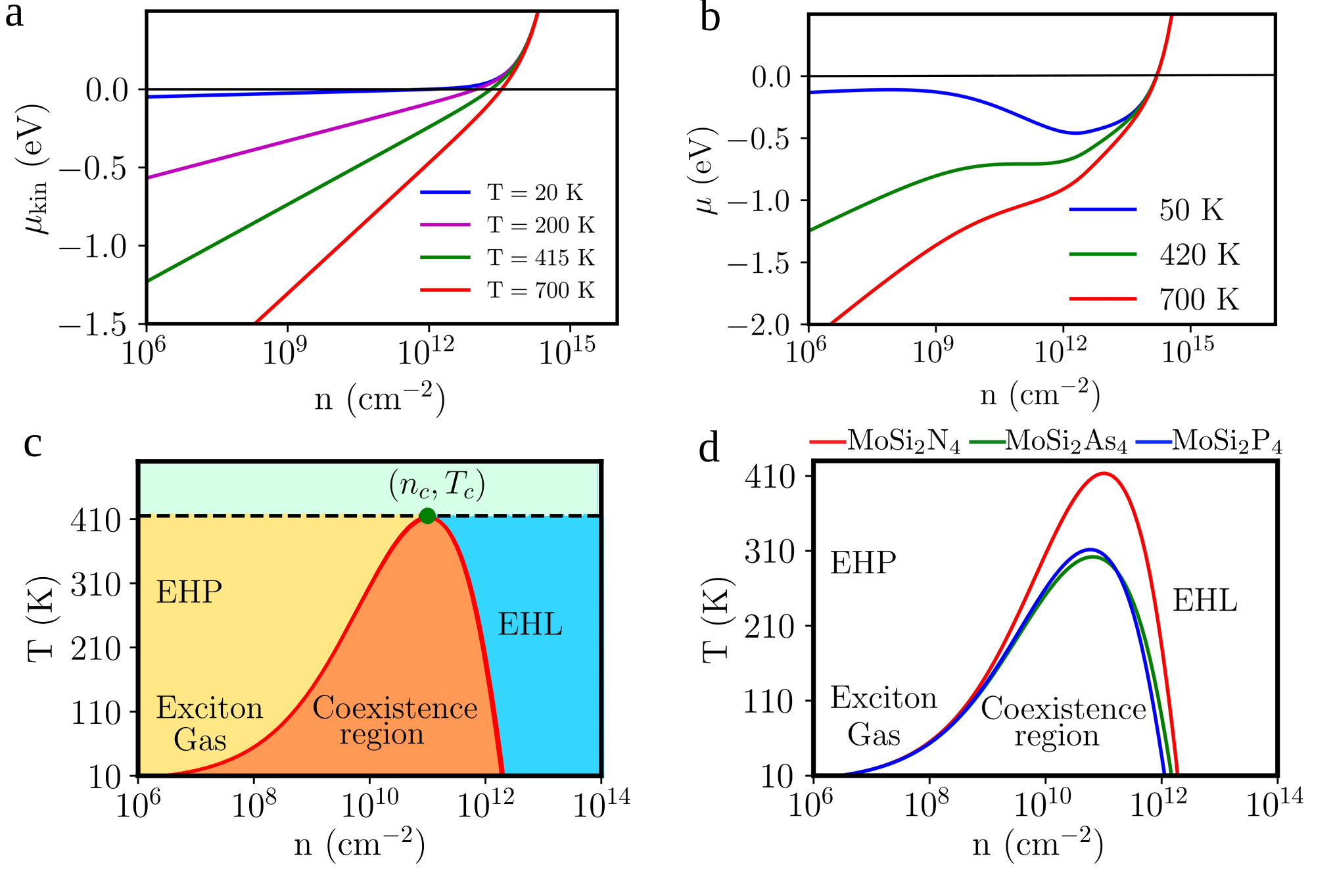}
	\caption {\textbf{a}) The density variation of the chemical potential contribution from the kinetic energy term at different temperatures. 
    \textbf{b}) The total chemical potential for the interacting electron-hole system. On including the exchange and correlation terms, the chemical potential becomes a non-monotonic function of density. Beyond a critical temperature $(T_c)$, the system can have two different densities at the same $\mu$, indicating the coexistence of the electron-hole gas and the condensed electron-hole liquid phase. \textbf{c}) The thermodynamic phase diagram in the $n-T$ plane shows regions of coexistence of an electron-hole gas and electron-hole liquid phase. 
    \textbf{d}) The phase boundaries and the coexistence region for all three {\formula} monolayers in the $n-T$ plane. 
	\label{Fig4}}
\end{figure*}

Together, the kinetic and exchange energy contributions form the Hartree-Fock energy. The remaining correlation energy contributions from the 
Coulomb interaction includes the second and higher order terms of the perturbation series~\cite{Combescot_1972, Rice1978-book}. 
These are typically captured by the Random phase approximation (RPA)~\cite{Combescot_1972}, which sums the infinite series of the bubble diagrams. Within the RPA approximation, the correlation energy has the following form,
\bea
E_{\rm{corr}} = \sum_{q}\int_{0}^{\infty} \frac{\hbar d\omega}{2\pi}\left[ \tan^{-1}\left (\frac{-B_q(\omega)}{1-A_q(\omega)}\right )+B_q(\omega) \right].
\label{eq4}
\eea
Here, $A_q(\omega)$ and $B_q(\omega)$ are the frequency-dependent real and imaginary parts of $V_q\chi(q,\omega)$, or 
$A_q(\omega) + iB_q(\omega) = V_q\chi(q,\omega)$.
$\chi(q,\omega)$ is the sum of the Lindhard susceptibility of the electrons and holes~\cite{Giuliani-Giovanni-Vignale}, and $V_q$ is the Keldysh potential defined in Eq.~\ref{eq3}. 
To understand the role of dimensionality in interaction effects, we calculate the correlation energies for different $r_0$ values. We find that a smaller $r_0$ value in the Keldysh potential yields the known correlation energy for the 2D case with unscreened Coulomb interactions. On increasing the $r_0$ value, the correlation energy decreases. This highlights that correlation effects become more pronounced on reducing the dimensions of the systems (Fig.~S3 of the SM \cite{Note1}).

Our numerical calculations of the total energy show that for low densities of the photo-excited carriers, both the exchange and correlation energies dominate over the kinetic energy contribution. 
With a gradual increase in the photo-excited carrier density, the kinetic energy dominates the correlation and exchange terms. This leads to a non-monotonic behavior in the total energy curve with a minimum at the equilibrium density [see Fig.~\ref{Fig3} (a)] for monolayer {\formulaN}. We find that the ground state energy for monolayer {\formulaN} is larger in magnitude than the ground state energies of {\formulaAs} and {\formulaP} [see Fig.~\ref{Fig3} (b)]. The relatively smaller effective thickness (or $r_0$) of {\formulaN} compared to {\formulaAs} and {\formulaP} make the exchange-correlation effects stronger, lowering its ground state energy. We show below that this makes the EHL phase in monolayer {\formulaN} relatively more stable with a higher $T_c$.   

\begin{table*}[t]
	\def \hsp{\hspace{10cm}}
	\caption{{The free exciton binding energy ($E_{\rm{x}}$), critical temperature ($T_c$), and the corresponding critical electron-hole pair density ($n_c$) from our calculations.}}	
	\vspace{0.15 cm}
	\begin{tabular}{c c c c}
		\hline \hline \vspace{1 mm}
		Compound \hsp    &  $E_x$ (eV) \hsp   & $T_c$ (K)  \hsp   &  $n_c$ ($\times 10^{11}$ $\mathrm{cm^{-2}})$ \\
		\hline
		\formulaN &  0.432 & 415 & 1.0 \\
		\formulaAs & 0.466 &  302 & 0.7 \\
		\formulaP  & 0.433 &  312 & 0.6 \\
		\hline
	\end{tabular}
	\label{table:2}
\end{table*}

The EHL droplet formation is a first-order phase transition similar to the gas-liquid transition. The EHL droplet condensation happens when a supersaturated electron-hole system exhibits a high-density liquid phase and a low-density gaseous phase simultaneously. This occurs at a critical density ($n_c$) and a critical temperature ($T_c$). Above the critical density and below the critical temperature the excitons lose their individuality, and electrons and holes condense into a droplet~\cite{Simon-2002}. To study the thermodynamics of the photo-excited electron-hole pairs, we calculate the free energy and derive the  chemical potential. 

The free energy per particle can be expressed as 
$
F (n, T) = F_0(n,T) + F_{\mathrm{xc}}(n, T).
$
Here, $F_0(n,T)$ is the free energy of the non-interacting electrons and holes, and $F_{\mathrm{xc}}(n,T)$ is the Coulomb interaction induced exchange and correlation contribution. At a high density of electrons and holes, the EHL is known to be metallic~\cite{Brinkman-72, Thomas-73,thomas-PRL-1974}. In the metallic regime, we can safely ignore the explicit $T$ dependence of the interaction part of the free energy and express $F_{\mathrm{xc}}(n, T) = F_{\mathrm{xc}}(n)$. 
Accordingly, the chemical potential is given by
\bea
\mu = \left( \frac{\partial F}{\partial N}\right)_{T,V} = \mu_{\rm{kin}} + \left( \frac{\partial F_{\rm{xc}}}{\partial N}\right )_{T,V}~.
\label{eq7}
\eea
The first term on the right-hand side of Eq.~\ref{eq7} refers to kinetic energy contribution to the chemical potential, while the second term refers to the contribution of the exchange and correlation energy to the chemical potential ($\mu_{\rm xc}$). These can be calculated from ~\cite{thomas-PRL-1974}, 
\bea
\mu_{\rm{kin}} & = & \frac{1}{\beta}\left( \ln[e^{\beta E^e_F} -1] + \ln[e^{\beta E^h_F} -1] \right)~, \nonumber \\ 
\mu_{\rm{xc}} & = & E_{\rm{xc}} + n \frac{\partial E_{\rm{xc}} }{\partial n}~.
\label{eq9}
\eea
Here, $\beta = 1/(k_BT)$ is the Boltzmann constant, and $E_{\rm{xc}}$ is the sum of exchange and correlation energies. 

We present the temperature and carrier density dependence of the $\mu_{\rm kin}$ for the monolayer {\formulaN} in Fig.~\ref{Fig4} (a). The dependence of $\mu_{\rm xc}$ on the photoexcited carrier density is presented in Fig.~S2 of the SM \cite{Note1}. The density dependent exchange-correlation chemical potential captures the renormalization of the bandgap and the corresponding exciton binding energy~\cite{BGR-Kalt-1992,MoS2-nano-lett} with changing density of photo-excited carriers (see Fig.~S2 and Sec.~S2 of the SM \cite{Note1} for details). From the $\mu_{\rm xc}(n)$ plot for monolayer {\formulaN}, we find that on including the effective thickness of the layer ($r_0$ = 99.15 \AA), the free exciton binding energy becomes 0.432 eV. Similarly, the free exciton binding energy calculated for the monolayer {\formulaAs} and {\formulaP} are summarized in Table~\ref{table:2}. These exciton binding energies in Table~\ref{table:1} are significantly lower than the exciton binding energies calculated from first principles which do not include the impact of effective layer thickness in the Coulomb interactions. 

The total chemical potential calculated using Eq.~\ref{eq7} is shown in Fig.~\ref{Fig4} (b). Depending on the temperature, there are regions with the possibility of having two different densities at the same chemical potential, a clear indication of the coexistence of two phases. For $T > T_c$, the chemical potential increases monotonically with exciton density. On reducing the temperature, we reach a critical temperature $T =T_c$, for which the slope of the chemical potential curve goes to zero at a critical density ($n = n_c$). This inflection point ($n_c, T_c$) marks the onset of the EHL phase transition. 
The boundary of the co-existence region of the liquid and the gas phase is determined by $\partial_{n} \mu = 0$ and the critical point is obtained from~\cite{Rice1978-book,Tikhodeev_1985}, or 
\bea
\frac{\partial \mu}{\partial n} \bigg |_{(n_c,T_c)} = \frac{\partial^2\mu}{\partial n^2} \bigg |_{(n_c,T_c)} = 0~.
\label{eq10}
\eea

We present the boundary of the coexistence region and the critical point for the {\formulaN} monolayer in the temperature-density plane in Fig.~\ref{Fig4} (c). The phase diagram clearly shows the `gas region' supporting free excitons and electron-hole plasma, the coexistence region, and the 
region with the electron-hole liquid. Our calculations suggest that 
the critical temperature $T_c = 415$ $K$, and the critical density $n_c = 1.0\times 10^{11}$ cm$^{-2}$  for monolayer {\formulaN}. Monolayers {\formulaAs} and {\formulaP} have a qualitatively similar phase diagram as shown in Fig.~\ref{Fig4}(d). The critical density and critical temperature for the EHL phase transition for all three materials is summarized in Table~\ref{table:2}.

\begin{figure}[t]
	\centering
	\includegraphics[width =\linewidth]{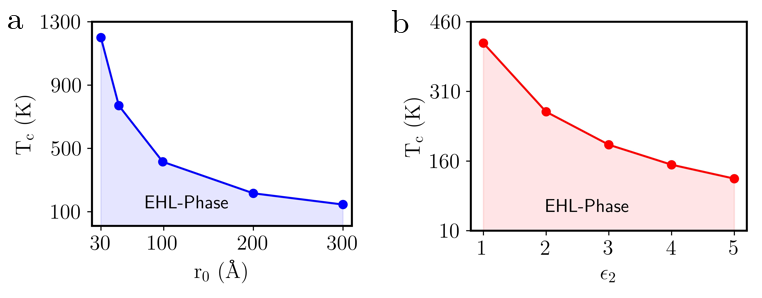}
	\caption {The impact of the effective thickness and the dielectric constant on the transition temperature of the EHL phase. \textbf{a}) The reduction of the EHL transition temperature with increasing $r_0$ and \textbf{b}) the $T_c$ variation with the dielectric constant of the substrate. Increase in both the parameters reduce the Coulomb interaction strength defined by Eq.~\ref{eq3}, and therefore it decreases the critical temperature. Broadly, anything that decreases the strength of the Coulomb interaction leads to a reduction in the $T_c$.
	\label{Fig5}}
\end{figure}

We find that all three monolayers can support room-temperature 
EHL phase. 
Amongst the three monolayers, {\formulaN} has the lowest $r_0$ and the highest $T_c$. A higher $r_0$ value decreases the strength of the effective Coulomb interactions and the exchange-correlation energy (see Fig.~{S3} of the SM \cite{Note1}) leading to a lowering of the $T_c$ which is demonstrated in Fig.~\ref{Fig5}~(a). More interestingly, we find that the critical density needed to achieve the EHL phase in all three monolayers is easily achievable in experiments \cite{amit-exciton-2}.  Amongst the three monolayers, the critical density for the EHL phase is lowest in {\formulaP}. This is a consequence of the lower (electron and hole) effective masses in {\formulaP} (see Table~\ref{table:1}). The qualitative criteria for EHL formation is $n > a_{ex}^{-2}$, where $a_{ex}$ is the exciton Bohr radius which depends on the exciton effective mass ($m^{*}$) and the dielectric constant of the material ($\epsilon$) as $a_{ex} \approx {\epsilon \hbar^2/(m^{*} e^2)}$. Here, $\hbar$ and $e$ are the reduced Planck's constant and electronic charge, respectively. This allows the electrons and holes in {\formulaP}  to condense into a macroscopic EHL phase at a relatively lower density. 

Figure~\ref{Fig4} demonstrates the possibility of room temperature EHL phase in the monolayers of the {\formula} series. However, our calculations rely on the specific choice of $r_0$ and the dielectric constant. For 2D materials, the substrate's dielectric constant can also significantly impact its optical properties and EHL phase [see Eq.~\ref{eq3}]. 
The increase in the effective dielectric constant ($\epsilon'$) decreases the strength of the Coulomb potential. This results in a reduction of the exchange and correlation energy or the magnitude of the total ground state energy of the electron-hole system with increasing dielectric constant. To quantify this variation, we show the dependence of the exchange energy, correlation energy, and the EHL phase boundary on the dielectric constant in Fig.~{S3} of the SM \cite{Note1}. As expected, an increasing the dielectric constant pushes the EHL phase boundaries towards lower temperatures. This becomes even more evident in Fig.~\ref{Fig5} (b), which shows the decrease of the $T_c$ with increasing dielectric constant. However, even with a dielectric constant of the  substrate $\epsilon_2=5$, {\formulaN} has a $T_c$ of more than 100 K for the EHL phase. 

In summary, we predict the possibility of observing the EHL phase at room temperature in the monolayer {\formula} family. Our calculations show that monolayer {\formulaN}, {\formulaAs}, and {\formulaP} are capable of supporting a stable EHL phase at room temperature with easily achievable  photo-excited carrier densities. This is due to the more prominent exchange and correlation effects in 2D systems compared to 3D systems, which helps stabilize the EHL phase at higher temperatures.
This exciting possibility can be experimentally verified through photoluminescence experiments~\cite{MOS2-EHP-Bataller2019, MoS2-EHL-EXP-ACS-Nano-2019-Yu} or by photocurrent spectroscopy experiments \cite{Arp2019}. Our study motivates further exploration of the impact of magnetic field, strain and other perturbations on the quantum droplets of the electron-hole liquid 
 for ultra-high power photonic applications.  

\vspace{.7 cm}
{\bf Methods:-} To find the minimum energy relaxed crystal structure for our first principle calculations, we performed symmetry-protected ionic relaxation while keeping the shape and volume of the unit cell fixed. The ionic positions were allowed to evolve until the residual Hellmann-Feynman force per atom was less than 0.0001 eV/{\AA}. We observed around 8\% change in the Mo-Z bond length after the relaxation. 
Using the relaxed lattice parameters, we perform the first-principle electronic structure calculation. To prevent spurious interlayer interactions, we applied a 27 {\AA} vacuum along the $c$-axis. We used the generalized gradient approximation (GGA)~\cite{GGA-PBE} to  incorporate the exchange and correlation effects. We used an  energy cutoff of 50 Ry for the plane-wave basis set, after a convergence test. A tolerance of $10^{-7}$ eV is used for electronic energy minimization. We used a $\Gamma$-centered $12\times 12\times 1$ Monkhorst k-mesh to perform the Brillouin zone integration~\cite{monkhorst}. All density functional theory-based calculations were performed using the Quantum ESPRESSO simulation package~\cite{QE,GGA-PBE}.

{For optical properties, we first calculate the quasiparticle (QP) self-energy starting from the electronic ground states calculated using GGA. We calculate the self energies for the electrons  and holes with a self-consistent GW method on eigenvalues only (evGW)~\cite{Hedin-GW1,Hybertsen-GW2,DFT_GWA_Exciton-GW3}. To evaluate the diagonal elements of the exchange self-energy, we used $10^6$ random points in our calculation with an energy cutoff of 50 Ry after a convergence test. To calculate the polarization function within the random-phase approximation, we have used three hundred forty bands and an energy cutoff of 14 Ry after a convergence test.}

\begin{acknowledgments}
A. A. acknowledges the Science and Engineering Research Board for Project No. MTR/2019/001520, and the Department of Science and Technology for Project No. DST/NM/TUE/QM-6/2019(G)-IIT Kanpur, of the Government of India, for financial support. We acknowledge the high-performance computing facility at IIT Kanpur for computational support. We also acknowledge the National Supercomputing Mission (NSM) for providing computing resources of `PARAM Sanganak' at IIT Kanpur, which is implemented by C-DAC and supported by the Ministry of Electronics and Information Technology (MeitY) and Department of Science and Technology (DST), Government of India. P. Y. acknowledges the UGC for Senior Research Fellowship.
\end{acknowledgments}
\bibliography{EHL_MoSi2Z4}
\end{document}


\title{Supplementary Material \par
\vspace{5mm} 
Room temperature electron-hole liquid phase in  
monolayer MoSi$_2$Z$_4$ (Z = pinctogen)}
\author{Pushpendra Yadav}
\email{pyadav@iitk.ac.in}
\affiliation{Department of Physics, Indian Institute of Technology Kanpur, Kanpur-208016, India}
\author{K. V. Adarsh}
\email{adarsh@iiserb.ac.in}
\affiliation{Department of Physics, Indian Institute of Science Education and Research Bhopal, Bhopal 462066, India}
\author{Amit Agarwal}
\email{amitag@iitk.ac.in}
\affiliation{Department of Physics, Indian Institute of Technology Kanpur, Kanpur-208016, India}

\maketitle

\begin{figure}
	\centering
	\includegraphics[width =.8\linewidth]{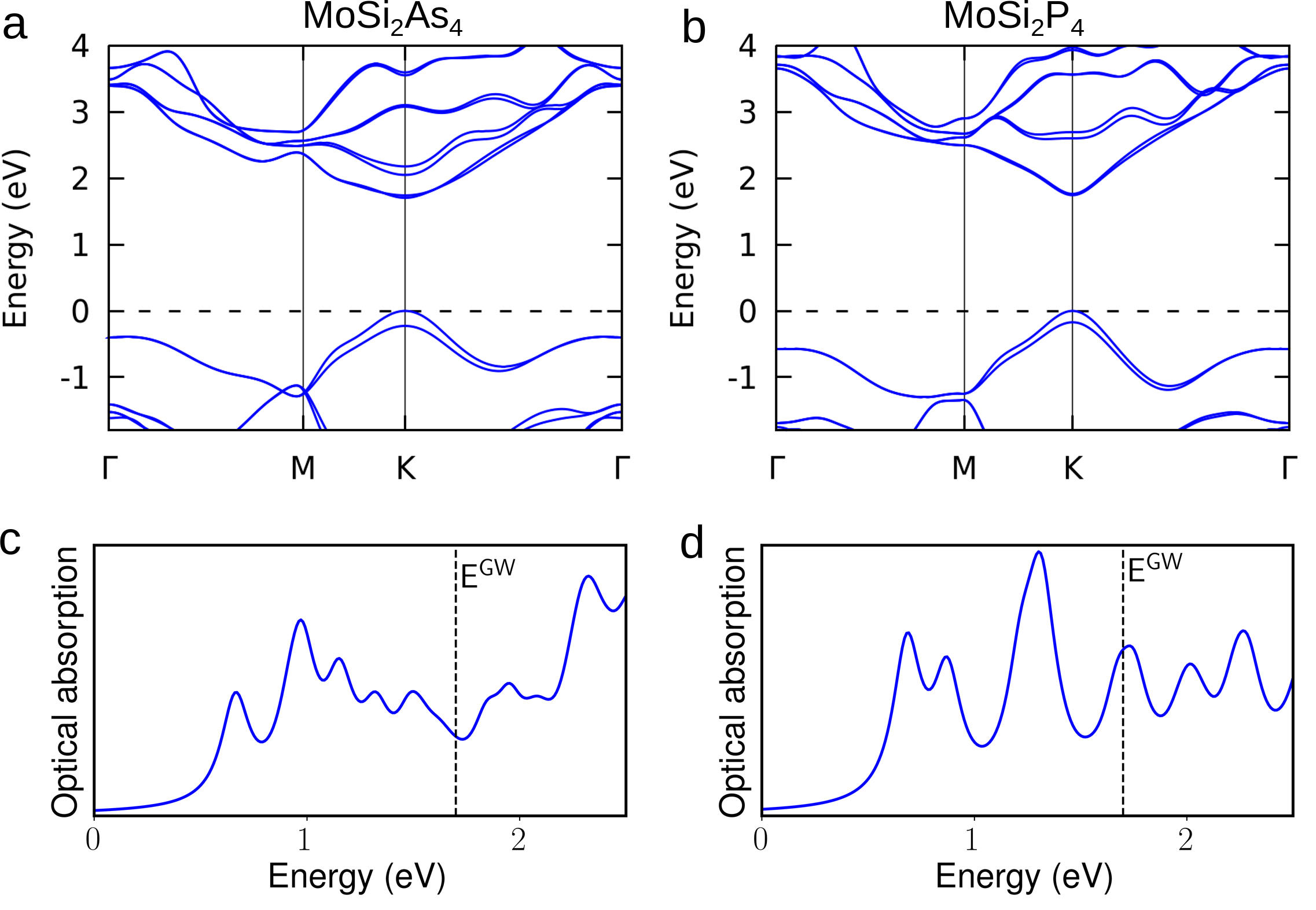}
	\caption {The quasiparticle bandstructure for monolayers of \textbf{a}) {\formulaAs} and  \textbf{b}) {\formulaP} calculated with spin-orbit coupling using evGW. The optical absorption spectra for each monolayer, calculated using the Bethe-Salpeter equation on top of the evGW bandstructure with spin-orbit coupling, are displayed in \textbf{c}) and \textbf{d}), respectively. The dashed vertical lines indicate the  evGW bandgap. The BSE optical spectrum captures excitonic resonances, manifesting as prominent absorption peaks below the evGW bandgap, and includes two-particle electron-hole interactions.}
\label{sm-fig1}
\end{figure}
\section{Quasiparticle bandstructure and optical absorption}
\label{sec-sm1}
To understand the optical excitations in monolayer {\formula} series, we begin by calculating the quasiparticle (QP) self-energy using the electronic ground states calculated using the generalized gradient approximation (GGA). We use the self-consistent GW method on eigenvalues only (evGW) to calculate the self-energies for electrons and holes. The QP band structure for the monolayer {\formulaAs} and {\formulaP} are shown in Fig.~\ref{sm-fig1} \textbf{a} and Fig.~\ref{sm-fig1} \textbf{b}, respectively. 
 
Further, we use the Bethe-Salpeter equation (BSE) to incorporate electron-hole interactions as discussed in Sec. II of the main text. By numerically solving the BSE, we extract the imaginary part of the dielectric function, which gives us the optical absorption spectrum. We perform quantum many-body perturbation theory calculations using the implementation in the YAMBO package. Our results indicate that monolayer {\formulaN} has an indirect bandgap of 3.58 eV at the evGW level, while {\formulaAs} and {\formulaP} have direct bandgaps of 1.70 eV and 1.74 eV, respectively. Importantly, these synthetic monolayers exhibit strong excitonic behavior as shown in Fig. 1 \textbf{c} and Fig. 1 \textbf{d}. The lowest exciton peak has a binding energy of approximately 1.35 eV, 1.04 eV, and 1.06 eV for monolayer {\formulaN}, {\formulaAs}, and {\formulaP}, respectively. 

\section{Binding energy and bandgap renormalization}
\label{section-sm2}
\begin{figure}
	\centering
	\includegraphics[width =.8\linewidth]{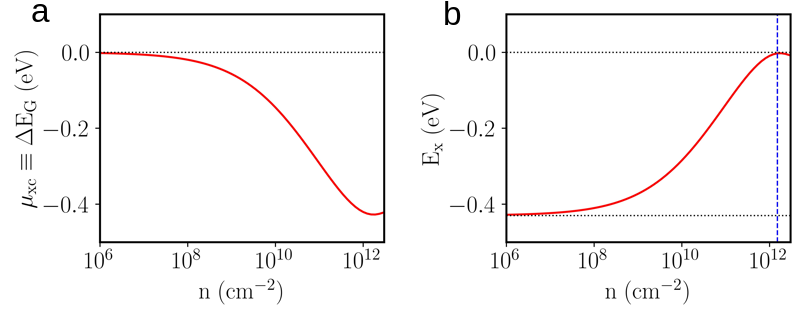}
	\caption {\textbf{a}) The bandgap renormalization as a function of the photo-excited carrier density $n$. \textbf{b}) The corresponding renormalized exciton binding energy as a function of $n$. The density at which the excitons are no more bound pair of electron and holes (with binding energy zero) is referred to as the Mott density. For monolayer {\formulaN}, we find the Mott density to be 1.5 $\times$ 10$^{12}$ cm$^{-2}$ and it is marked by the blue dotted vertical line in panel \textbf{b}).
\label{sm-fig2}}
\end{figure}

One important aspect of the variation in the exciton binding energy with photoexcited carrier density is the band gap renormalization (BGR). BGR captures the impact of the correlation and exchange energy in reducing the single particle energy bandgap. The variation in the chemical potential induced by the exchange and correlation energy terms gives an approximate estimation of the BGR as a function of the number density of photoexcited electron-hole pairs. We have calculated the change in the bandgap, $\Delta E_g \approx \mu_{xc}$, where $\mu_{xc}$ is defined by Eq.~6 in the main text~\cite{MoS2-nano-lett,BGR-Kalt-1992}. We present the variation of the exchange correlation part of the chemical potential and the BGR in {\formulaN} in Fig.~\ref{sm-fig2}. We find that beyond a critical density of photoexcited carriers (referred to as the Mott density), the excitons merge with the continuum single-particle excitations. 
For \formulaN, we find the Mott density to be approximately 1.5 $\times 10^{12}$ cm$^{-2}$.

\section{Impact of layer thickness and dielectric constant on the EHL phase}
\label{section-sm3}

\begin{figure}
	\centering
	\includegraphics[width =.9\linewidth]{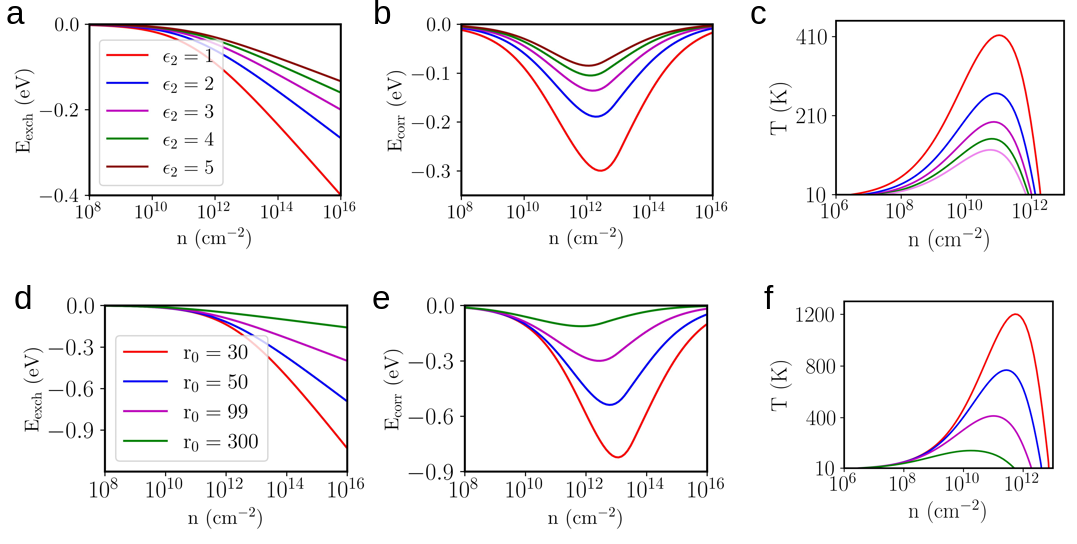}
	\caption {The sensitivity of the EHL transition temperature on the dielectric constant and the effective thickness. The variation of the \textbf{a}) exchange energy, \textbf{b}) correlation energy, and \textbf{c}) the phase boundary of monolayer {\formulaN} with varying dielectric constant of the substrate. Similarly, the effect of the effective layer thickness $r_0$ on the  \textbf{d}) exchange energy,  \textbf{e}) correlation energy, and  \textbf{f}) the phase diagram of the monolayer {\formulaN}.
 \label{sm-fig3}}
\end{figure}
Figure~4 of the main text demonstrates the possibility of room temperature EHL phase in the monolayers of the {\formula} series. Here, we show the dependence of the exchange energy, correlation energy, and the EHL phase boundary on the dielectric constant in Fig.~\ref{sm-fig3} \textbf{a-c}. 
Increasing the dielectric constant reduces the strength of the Coulomb interactions. This pushes the EHL phase boundaries towards lower temperatures. Figure 5~\textbf{a} in the main text shows that the $T_c$ decrease with increasing dielectric constant. 
 We also check the sensitivity of the $T_c$ with the effective thickness, $r_0$, in the Keldysh potential defined in Eq. 3 of the main text. $r_0$ captures the dimensional crossover of the Coulomb potential from 2D to an effective 3D. Increasing $r_0$  decreases the strength of the Coulomb interaction and the $T_c$ of the EHL phase. To observe the quantitative impact of $r_0$, we show the variation of the exchange and correlation energies of the electron-hole system with increasing $r_0$ in Fig.~3 \textbf{d} and Fig.~3 \textbf{e}. The exchange and correlation energies increase in magnitude as $r_0$ decreases. This reflects in the lowering of the $T_c$ with increasing $r_0$, as shown in Fig.~5 \textbf{d} and a quantitative variation of $T_c$ with $r_0$ is shown in Figure 5 (\textbf{b}) of the main text. 
 
\bibliography{ref_EHL}